\documentclass[12pt,twoside]{article}   
\usepackage[super,sort,comma]{natbib}

\usepackage{comment}
\usepackage{fancyhdr}		

\newcommand{\captionv}[3]{\begin{center}\parbox{#1cm}{\caption[#2]{{\sf #3}}}
        \end{center}}

\usepackage[section]{placeins}   %

\usepackage{graphicx}
\usepackage{multirow}
\usepackage{booktabs}
\usepackage{tabularx}    
\usepackage{array} 

\makeatletter \renewcommand\@biblabel[1]{$^{#1}$} \makeatother
\usepackage[mathlines]{lineno}

\usepackage{hyperref}
\hypersetup{ colorlinks,
    citecolor=blue,
    filecolor=blue,
    linkcolor=blue,
    urlcolor=blue
}

\usepackage{xcolor}

\definecolor{gray}{rgb}{0.6,0.6,0.6}
\definecolor{red}{rgb}{0.85,0,0}
\definecolor{green}{rgb}{0,0.85,0}
\definecolor{blue}{rgb}{0,0,0.85}
\definecolor{beige}{rgb}{0.92,0.87,0.78}
\usepackage[all]{hypcap}    
\usepackage{amsmath}
\usepackage{amsfonts}
\usepackage{amssymb}
\usepackage{algorithm2e}
\usepackage{mathrsfs}
\usepackage{enumitem}

\newcommand{\rev}{\color{black}}

\begin{document}

\begin{center}
\sf{\Large {\bfseries Investigating the Uncertainty of Cellular Microenvironment Parameter Estimations via Diffusion MRI Cytometry}} \\  
\vspace*{10mm}
Wen Li$^1$, Yan Dai$^2$, Arely Perez Rodriguez$^2$, Todd Aguilera$^2$, Jie Deng$^2$, Xun Jia$^1$ \\
\vspace*{10mm}
$^1$Department of Radiation Oncology and Molecular Radiation Sciences, Johns Hopkins University, Baltimore, MD, USA

$^2$Department of Radiation Oncology, University of Texas Southwestern Medical Center, Dallas, TX, USA\\
\end{center}
\pagenumbering{roman}
\setcounter{page}{1}
\pagestyle{plain}

\begin{abstract}
\noindent {\bf Background:}
Cell microenvironment characteristics serve as critical biomarkers for early tumor response assessment in radiation therapy. Diffusion MRI (dMRI) provides a non-invasive approach to probe these microenvironment features; however, conventional model-fitting approaches for microenvironment parameter estimation often suffer from high uncertainty and poor robustness.

\noindent {\bf Purpose:}
This study aims to establish a theoretical foundation for identifying cell microenvironment parameters that can be robustly estimated from signals of IMPULSED dMRI and to develop a reliable mapping-based estimation framework for these parameters.

\noindent {\bf Methods:}
We simulated dMRI signals using the well-established IMPULSED model, incorporating a pulsed gradient spin echo (PGSE) sequence and two oscillating gradient spin echo (OGSE) sequences with two different frequencies. The simulations were based on five cellular parameters: cell diameter ($d$), intracellular diffusion coefficient ( $D_{in}$), intracellular volume fraction ( $V_{in}$), extracellular diffusion coefficient ( $D_{ex}$), and the slope of $D_{ex}$ with respect to oscillation frequency ($\beta_{ex}$). Parameter uncertainty was quantified using Jacobian-based sensitivity analysis at a signal-to-noise ratio (SNR) of 30, corresponding to achievable clinical conditions on a 1.5T MRI scanner. To develop direct parameter mapping models, dMRI signals were logarithmically transformed to enhance linearity and reduced in dimension using principal component analysis (PCA). The logarithm-transformed dimension-reduced data were used to estimate microenvironment parameters using three mapping models: linear regression, fourth-order polynomial regression, and a fully connected 4-layer neural network. Model validation was further performed in an \textit{in vitro} experiment using MC38 cell lines.

\noindent {\bf Results:}
Uncertainty analysis identified $d$,  $V_{in}$, and  $D_{ex}$ as robustly derivable parameters, each with relative uncertainty below 1.0. Among the models considered, the 4-layer neural network achieved the best performance, yielding mean absolute errors (MAE) of 1.7~$\mu$m for $d$, 5.06\% for  $V_{in}$, and 0.28~$\mu$m$^2$/ms for  $D_{ex}$. In the \textit{in vitro} experiment, the model achieved a 6.7\% error in cell diameter estimation.

\noindent {\bf Conclusions:}
This study identified the cell microenvironment parameters that can be robustly estimated from IMPULSED dMRI signals and established a mapping-based framework for accurate and robust parameter estimation. The proposed approach provided a foundation for non-invasive, quantitative assessment of tumor microenvironment changes for monitoring radiation therapy response.

\end{abstract}

\newpage     

\tableofcontents

\newpage

\setlength{\baselineskip}{0.7cm}      

\pagenumbering{arabic}
\setcounter{page}{1}
\pagestyle{fancy}

\section{Introduction}
Early tumor response assessment is critical for improving cancer radiotherapy outcomes. By evaluating  tumor response to radiation at an early stage, it becomes possible to rapidly identify non-responders or patients with suboptimal responses, enabling timely adjustments to the initial treatment plan and thereby achieving personalized, precise radiotherapy\citep{therasse2000new}. Detailed information about the cell microenvironment, such as cell size and intracellular volume fraction, provides direct and essential insights into the biological changes induced by treatment and can serve as valuable biomarkers for early response assessment. Traditionally, biopsy is considered as the gold standard for acquiring such microenvironment information\citep{newton2011progress}. However, biopsies are invasive, time-consuming, and not always feasible for repeated monitoring. In recent years,  Diffusion-weighted imaging offers a safe and non-invasive alternative to probe tumor microenvironment\citep{swartz2018influence,bai2021study,jordan2005dynamic}. The principle of this diffusion MRI (dMRI)-based cell microenvironment analysis method, also termed MRI cytometry\citep{xu2021mri}, is based on the random motion of water molecules within tissues. Since water diffusion is hindered by cellular structures, dMRI signals can reflect alterations in the microenvironment\citep{sigmund2011intravoxel}. Because treatment responses are initially manifested as cellular-level changes before becoming evident as macroscopic symptoms, probing cell microenvironment with dMRI provides a valuable method for early tumor response assessment\citep{chenevert1997monitoring}.

MRI cytometry is typically achieved by measuring signals obtained using combined pulsed gradient spin echo (PGSE) and oscillating gradient spin echo (OGSE) sequences. Conventional PGSE sequence is straightforward to apply, but relatively insensitive to short distance cellular scales. In contrast, OGSE sequences, particularly at high frequencies, can probe shorter diffusion times (less than 5 $\mu$m), allowing for the detection of microenvironment changes  over short length scales \citep{xu2009quantitative}, but faces challenges when assessing relatively large cell sizes (greater than 10 $\mu$m) due to limitations in the maximum achievable diffusion time within practical echo times\citep{jiang2016quantification}. In practice, a hybrid approach that  acquires signals with both sequences has been demonstrated to achieve accurate estimations for cell sizes and intracellular volume fractions \textit{in vitro} for cell diameters ranging from 10 to 20 $\mu$m, suitable for assessing microenvironment changes associated with anticancer treatment \citep{jiang2016quantification}. 

Based on specific sequences, theoretical models have been derived to relate the dMRI signal with parameters describing cell microenvironment, e.g. cell diameter ($d$), intracellular diffusion coefficient ($D_{in}$), intracellular volume fraction ($V_{in}$), extracellular diffusion coefficient ($D_{ex}$), and the slope of extracellular diffusion coefficient with respect to oscillation frequency ($\beta_{ex}$)) \citep{stejskal1965use,callaghan1993principles,cory1990measurement,brownstein1979importance,tanner1968restricted,yablonskiy2003statistical,zhao2008intracellular}. Based on a theoretical model, cell microenvironment parameters can be estimated from measured dMRI signal via data fitting. A certain loss function is minimized to derive cell microenvironment parameters that yields a dMRI signal matching the measured one under the theoretical model. While conceptually straightforward,  due to the highly nonlinear and complex signal model in MRI cytometry, the noise in the measured dMRI signal is transferred to, and often amplified in the estimated cell microenvironment parameters during the data fitting process. Mami et al. \citep{iima2014characterization} highlighted that bi-exponential model fitting is highly sensitive to noise and the initial values. In contrast, parameter estimation using the kurtosis model, which has one fewer degree of freedom was more robust but lost accuracy when high $b$-values (greater than 2000-2500 s/mm$^2$) were used. Ades-Aron et al. \citep{ichikawa2018histological} compared different fitting methods and further demonstrated that microenvironment parameter estimates are strongly dependent on the fitting method employed. To improve parameter estimation accuracy, Ichikawa et al. \citep{ades2018evaluation} applied MP-PCA denoising and Rician bias correction to reduce dMRI noise prior to model fitting. Although this approach yielded some improvements, the results remained sub-optimal. Xu and Oeschger et al. \citep{xu2021mri,oeschger2023axisymmetric} investigated the impact of varying Signal-to-Noise-Ratio (SNR) levels on parameter estimation in MRI cytometry and found that lower SNRs significantly increased prediction variability. To address this, they performed multiple fittings (hundreds of times) using different noise but at the same SNR level, resulting in accurate average estimates. However, such an approach is not desired in clinical settings. 

For clinical practice, a robust and reliable cell microenvironment parameter estimation method is essential. Considering the high sensitivity of model-based fitting methods to noise effect \citep{iima2021perfusion}, it is critical to understand the mathematical properties of the dMRI signal model for developing accurate and robust models for estimating cellular microenvironment parameters. Such understanding provides theoretical insights into the inversion process from dMRI signals to estimated parameters, and informs the design of numerical methods that yield stable and reliable estimates. This is especially important considering that clinical translation in radiotherapy often relies on 1.5 T MRI scanners available at radiotherapy clinics, which, compared to the commonly used 3.0 T scanners in diagnostic settings, further amplifies measurement noise. 

Toward this goal, in this study, we specifically investigate the properties of a representative dMRI-based MRI cytometry model,  Imaging Microenvironment Parameters Using Limited Spectrally Edited Diffusion (IMPULSED) \citep{xu2020magnetic}, with a focus on its robustness to noise. We comprehensively analyze the numerical properties of the signal model to characterize how measurement noise propagates through the inversion process, and we derive theoretical guidance to inform robust parameter estimation strategies. 

The second contribution of this study is a novel estimation framework that couples a dimension reduction technique with a mapping model. Instead of directly fitting the full signal model to noisy measurements, a process that is often unstable and sensitive to noise, we derive a stable mapping from measurements to the estimated underlying microenvironment parameters using machine learning-based approaches, given the well-demonstrated effectiveness of machine learning in various medical imaging and radiotherapy tasks\citep{shen2020introduction,li2022multi}.
We systematically investigated different options for dimension reduction settings and mapping models to identify the optimal one. We validated our approach in preliminary \textit{in vitro} experiments using data acquired from a 1.5 T MRI scanner, demonstrating its potential feasibility in clinically realistic settings.



\section{Methods}

\subsection{dMRI signal model and data generation}

We focused on the IMPULSED model that describes the microenvironment with intra and extracellular components\citep{jiang2016quantification}. In this model, tissues are modeled as densely packed, spherical cells with an average cell diameter $d$, and the water exchange between the intracellular and extracellular spaces is neglected (i.e., impermeable) due to the limited impact to dMRI signal, as suggested by Assaf et al.\citep{assaf2004new}. The efficacy of this model has been extensively demonstrated in previous studies\citep{alexander2010orientationally,xu2014mapping,assaf2004new,jiang2016quantification,li2014fast}.

In the IMPULSED approach, the dMRI signal can be analytically expressed as weighted combination of the intracellular signal magnitude $S_{in}$ and extracellular signal magnitude $S_{ex}$: 
\begin{equation}
    S(b) = V_{in} \cdot S_{in}(b) + (1 - V_{in}) \cdot S_{ex}(b),
    \label{Eq_S}
\end{equation}
where $S$ represents the dMRI signal and $V_{in}$ is the water fraction of the intracellular space. A typical experiment include both PGSE and OGSE sequences \citep{jiang2016quantification} to employ PGSE for signals with long diffusion times, and OGSE with two different frequencies to acquire signals with shorter diffusion times. 

The OGSE signal is measured using cosine-modulated gradient waveforms. The intra and extracellular signal can be expressed as:
\begin{equation}
    \begin{split}
S_{in} (b|\text{OGSE}) &= \exp \Bigg( 
-2 (\gamma G)^2 \sum_{k} \frac{B_k \lambda_k^2 D_{in}^2}{(\lambda_k^2 D_{in}^2 + 4\pi^2 f^2)^2} 
\Bigg\{ \frac{(\lambda_k^2 D_{in}^2 + 4\pi^2 f^2)}{\lambda_k D_{in}} 
\Bigg[ \frac{\delta}{2} + \frac{\sin(4\pi f \delta)}{8\pi f} \Bigg] \\
&\quad - 1 + \exp(-\lambda_k D_{in} \delta) 
+ \exp(-\lambda_k D_{in} \Delta) (1 - \cosh(\lambda_k D_{in} \delta)) 
\Bigg\} \Bigg),\\
S_{ex}(b|OGSE) &= \exp\left[-b\left(D_{ex} + \beta_{ex} \cdot f\right)\right],
    \end{split}
    \label{Eq_Sin_OGSE}
\end{equation}
where $B_k$ and $\lambda_k$ are parameters that depend on the cell diameter. $\gamma$ denotes the gyromagnetic ratio, $G$ is the diffusion gradient strength,  $D_{in}$ is the intracellular free diffusion coefficient, and $f$ is the oscillation frequency. $\delta$ is the gradient duration and $\Delta$ represents the gradient separation. $b$ is the diffusion weight $b = \gamma^2 \int_0^{+\infty} dt |\int_0^t dt' g(t'|G, \delta, \Delta)|$, $g(\cdot)$ is the diffusion pulse's amplitude as a function of time. For the extracellular signal,  $D_{ex}$ is a constant, and  $\beta_{ex}$ denotes the slope of the extracellular free diffusion coefficient with respect to the oscillation frequency $f$. Here,  $D_{ex}$ is mainly influenced by extracellular tortuosity, while  $\beta_{ex}$ indirectly reflects microenvironment properties.


When $f \to 0$, the consine-modulated OGSE pulse degenerates into a PGSE pulse, and the intracellular and extracellular signals can be expressed as:
\begin{equation}
    \begin{split}
        S_{in}(b|PGSE) &= \exp \Bigg( 
        -2 \left(\frac{\gamma G}{D_{in}}\right)^2 
        \sum_{k} \frac{B_k}{\lambda_k^2} 
        \Bigg[ \lambda_k D_{in} \delta - 1 
        + \exp(-\lambda_k D_{in} \delta) \\
        &\quad + \exp(-\lambda_k D_{in} \Delta) 
        \left(1 - \cosh(\lambda_k D_{in} \delta) \right) 
        \Bigg] \Bigg),\\
        S_{ex}(b|PGSE) &= \exp(-bD_{ex}).
    \end{split}
    \label{Eq_Sin_PGSE}
\end{equation}


With these signal models, we generated data of the dMRI signals for a set of cell microenvironment parameters ($d$, $D_{in}$, $V_{in}$, $D_{ex}$, $\beta_{ex}$). These parameters were chosen to cover clinically relevant ranges of $d$ $\in$ [6.0 - 20.0]~$\mu$m, $D_{in}$, $D_{ex}$ $\in$ [0.2 $\mu$m$^2$/ms - 3.38 $\mu$m$^2$/ms]), $V_{in}$ $\in$ [$0\% - 100\%$], $\beta_{ex}$ $\in$ [0 $\mu$m$^2$ - 10 $\mu$m$^2$])\citep{xu2021mri}. We considered an experimental setup with a PGSE sequence with seven $b$ values and two OGSE sequences (OGSEn1 and OGSEn2) with cycle number $N=1,2$. The two OGSE sequences included five and four $b$ values, respectively \citep{xu2020magnetic}. The sequence parameters are listed in Table \ref{tab:pameters}. Together, for each microenvironment parameter, we computed 16 dMRI signal values at 16 $b$ values. 

\vspace{10pt}
\begin{table}[htbp]
    \centering
    \captionv{10}{}{Diffusion sequence parameters used in the study.}
    \label{tab:pameters}
    \renewcommand{\arraystretch}{1.2}
    \resizebox{1\linewidth}{!}{
    \begin{tabularx}{0.9\textwidth}{c|c|c|c}
    \hline\hline
        Sequence & $\delta$ (ms) & $\Delta$ (ms) & $b$ ($10^{-3}$ ms/$\mu$m$^2$)\\
        \hline
         PGSE  & 21 & 68 & \multicolumn{1}{l}{0, 250, 500, 750, 1000, 1250, 1500}\\
         \hline
         OGSEn1  & 60 & 68 & \multicolumn{1}{l}{0, 300, 600, 900, 1200}\\
         \hline
         OGSEn2  & 60 & 68 & \multicolumn{1}{l}{0, 100, 200, 300}\\
         \hline\hline
    \end{tabularx}
    }
\end{table}
\vspace{10pt}

To simulate clinical realistic PGSE and OGSE signals, Rician noise was added using the formulation: 
\begin{equation}
    \ S_n(b) = |[S(b)\cos(\phi)+n_1]+\mathrm{i}[S(b)\sin(\phi)+n_2]|,
    \quad \text{where } n_1, n_2 \sim \mathcal{N}(0, \nu^2),
    \label{eq:noise}
\end{equation}
where $S(b)$ and $S_n(b)$ represent the noise-free and noise-added dMRI signals at a given $b$ value respectively, while $n_1$ and $n_2$ denote independent Gaussian noise components for the real and imaginary parts. $\nu$ is the noise level, and $\phi$ is the phase factor uniformly sampled in $[0,2\pi)$. In this study, we considered the scenario with SNR of $\sim 10-30$, a typical range achievable for our 1.5 T MRI scanner (Ingenia Ambition X, Philips Healthcare, Netherlands). This converts to $\nu \approx 0.033-0.10$. Once the noise signal was generated, we normalized the signals using signal value at $b = 0$, yielding the simulated dMRI signal considering noise.


\subsection{Quantifying uncertainty of estimated microenvironment parameters}

Let us generally denote the signal function described in the previous section, namely Eq.~(\ref{Eq_S}) as $\mathbf{S}=f(\mathbf{x})$, where $\mathbf{x} = [d, D_{in}, V_{in}, D_{ex}, \beta_{ex}]'$ is a vector of the microenvironment parameters, $\cdot '$ denotes vector transpose, and $\mathbf{S}$ is a vector with 16 elements corresponding to the noise-less signals under the PGSE and OGSE sequences. $f(\cdot)$ stands for the signal functions. To investigate the uncertainty in the estimated microenvironment parameters due to measurement noise, we began by linearizing the signal model. The relationship between small perturbations in the signal $\delta \mathbf{S}$ and the corresponding perturbations in the parameters $\delta \mathbf{x} $ is:
\begin{equation}
    \mathbf{\delta S} = \mathbf{J(x}) \mathbf{\delta x},
    \label{eq:expansion}
\end{equation}
where \( \mathbf{J(x}) \) is the Jacobian matrix of the signal model $f(\cdot)$ with respect to \( \mathbf{x} \). Because the complex analytical expression of $f(\cdot)$, we computed \( \mathbf{J(x}) \) numerically using a finite difference approximation, where each parameter in $\mathbf{x}$ was perturbed individually by a small value ($\epsilon = 10^{-5}$) to estimate the corresponding partial derivatives of the signal vector $\mathbf{S}=f(\mathbf{x})$. Specifically, the element in the Jacobian matrix is given by:
\begin{equation}
    \textbf{J}(\mathbf{x})[i,j]=
 \dfrac{f_i(\mathbf{x} + \epsilon\mathbf{e}_j) - f_i(\mathbf{x}) }{ \epsilon},
    \label{eq:Jacobian element}
\end{equation}
where $\textbf{J}[i,j]$ represents the partial derivative of the $i$-th output signal $\mathbf{S}$ with respect to the $j$-th input parameter in $\mathbf{x}$.  Here, $\mathbf{e}_j \in \mathbb{R}^{5}$ is the standard basis vector with 1 in the $j$-th entry and 0 elsewhere. 

As the Jacobian matrix depends on the parameter $\mathbf{x}$, we considered in our study sampling each of the five microenvironment parameters at five quantile points (10\%, 30\%, 50\%, 70\% and 90\%) across its respective range, resulting in $n=5^5=3125$ different parameters $\textbf{x}$, and we computed the corresponding $\mathbf{J(x)}$.

Assuming the measurement noise is zero-mean and uncorrelated with variance $\nu^2$, namely $\mathrm{E}[\mathbf{\delta S} \mathbf{\delta S}^{'} ]=\nu^2\mathbf{1}$, where $\mathbf{1}$ is the identity matrix, we estimated the covariance matrix of $ \mathbf{\delta x}$ as:
\begin{equation}
    \mathbf{C}(\mathbf{x}) = \mathrm{E}[\mathbf{\delta x} \mathbf{\delta x}'] = \mathrm{E}[\mathbf{J}^{+} \mathbf{\delta S} \mathbf{\delta S}^{'} \mathbf{J}^{+'}] = \nu^2 \mathbf{J}^{+} \mathbf{J}^{+'} = \nu^2[\mathbf{J(x)}^{'} \mathbf{J(x)}]^{-1},
    \label{eq:vcv}
\end{equation}
where $\mathbf{J(x)}^{+}$ denotes the Moore–Penrose pseudo-inverse of \( \mathbf{J(x)} \). The resulting covariance matrix $ \mathbf{C}(\mathbf{x}) $ captures the uncertainty in the estimated parameters: its diagonal elements represent the variances of individual parameters, while off-diagonal elements represent the covariances between them. Since we are interested in the uncertainty of individual parameter, we extracted the diagnoal elements of $\mathbf{C}(\mathbf{x})$ and computed the uncertainty estimation by taking the square root, denoted as $\sigma_i(\mathbf{x})$, with $i=1,\ldots,5$ indicating the five microenvironment parameters. Finally we computed the median of $\sigma_i(\mathbf{x})$ over the set of microenvironment parameter $\mathbf{x}$, yielding the uncertainty estimation $\sigma_i$ characterizing the uncertainty level of the $i$th microenvironment parameter over the range of interest.

To express the uncertainty in a scale-invariant and relative form, we further defined a normalization factor (13.0 $\mu$m, 1.79 $\mu$m$^2$/ms, 50\%, 1.79$\mu$m$^2$/ms, 5.0 $\mu$m$^2$), which included the mid-point value of the range of each microenvironment parameter. We computed the relative uncertainty $\delta_i$ by normalizing $\sigma_i$ with the corresponding normalization factor.

Finally, we imposed an empirically chosen threshold level $\delta_0 = 1.0$ for the relative uncertainty.  As will be demonstrated in the Results section, the relative uncertainty values $\delta_i$ across different parameters exhibit a natural gap. This specific threshold was therefore selected to effectively divide the variables into two distinct groups: those that are robustly estimated and those that are non-robust. For the parameter $i$ with $\delta_i\le \delta_0$, we considered the estimation of them as relatively robust, hence focusing on constructing models to estimate these parameters in the next subsection.

\subsection{Developing a robust microenvironment parameter estimation model} 
\label{sec:model develop}

For each parameter to be estimated, we would like to construct a direct mapping function from the measured dMRI signal $\mathbf{S}$ to the corresponding parameter value $\mathbf{x}$ using a data-driven approach. Compared to the conventional method, which estimates all five microenvironment parameters by fitting the signal model in Eq.~(\ref{Eq_S}) to the measured dMRI data\citep{iima2014characterization, ichikawa2018histological, ades2018evaluation, xu2021mri, oeschger2023axisymmetric, iima2021perfusion}, this strategy offers several advantages, which will be presented in the discussion section.



To construct such a mapping, we first generated a synthetic dataset using the signal model described previously. Specifically, we randomly sampled $10^5$ independent microenvironment parameter vectors \(\mathbf{x}\) within the predefined parameter ranges. For each sampled \(\mathbf{x}\), we computed 10 noisy dMRI signals based on Eq.~(\ref{eq:noise}), thereby incorporating realistic measurement variability. In total, this procedure yielded \(10^6\) paired samples of \(\mathbf{x}\) and corresponding dMRI signals, which were subsequently used for model development.

\vspace{20pt}
\begin{figure}[hb]
   \begin{center}
   \includegraphics[width=16.5cm]{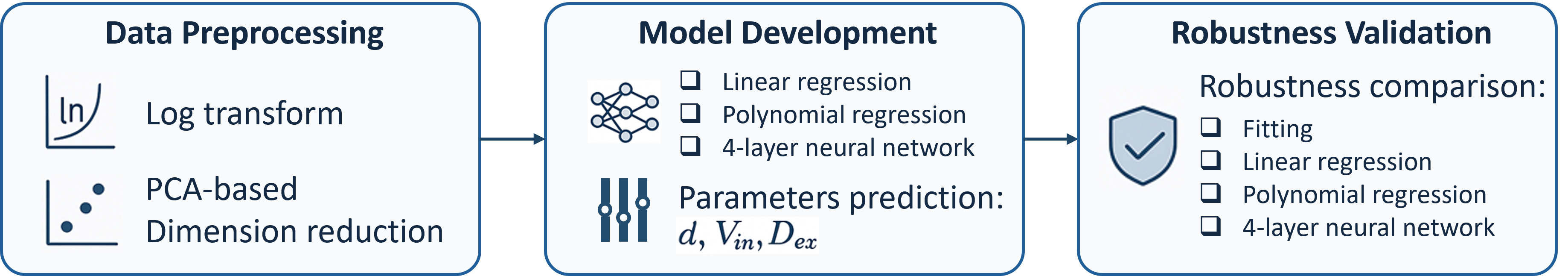}
   \captionv{12}{}{Workflow of model development for cell microenvironment parameters estimation.}
    \label{study_design} 
    \end{center}
\end{figure}

The overall workflow of the parameter  estimation model is shown in Fig.~\ref{study_design}. First, as the dMRI signal approximately follows an exponential decay form as a function of $b$, we applied a logarithmic transformation to the signal values to partially linearize the relationship and reduce the influence of the exponential decay. This preprocessing step facilitates effective subsequent dimensionality reduction and parameter mapping. 

Second, noticing that the dMRI signals corresponding to different microenvironment parameter sets $\mathbf{x}$ are generally similar, and that it is the relatively small variations in the signal with respect to $\mathbf{x}$ that are critical for parameter inference, we performed a dimension reduction step to capture the dominant modes of signal variation. In addition, high-dimensional data can lead to overly complex prediction models that risk fitting noise in the training data rather than the underlying physical patterns. Dimensionality reduction mitigates this issue by simplifying the input space, thereby improving the robustness and generalizability of the subsequent parameter estimation model. For simplicity, we employed Principal Component Analysis (PCA) as the dimension reduction technique. The dimension reduction of PGSE, OGSEn1 and OGSEn2 signals were performed separately.  For this step, a key question was the numbers of dimensions reduced to for each of PGSE, OGSEn1 and OGSEn2 signals. We exhaustively searched the possible combinations to identify the optimal choice.  

Finally, we established a function mapping from the logarithm-transformed dimension-reduced signal to the targeted cell parameters. In this step, three candidate mapping models were considered, including linear regression, fourth-order polynomial regression, and a neural network with 4 fully connected layers (labeled as 4-layer neural network). Given the relatively simple data fitting task, we empirically found this neural network was adequate. For the linear and polynomial regression models, model coefficients were computed using a least-squares solver. For the neural network, the synthesized dataset was randomly divided into training and validation subsets with a ratio of 7:3. LeakyReLU activation functions were applied after the first three fully connected layers. The network was trained under a supervised learning scheme by minimizing the L1 loss using the Adam optimizer with a fixed learning rate of 0.001 for 20000 epochs.

For each cell parameter, 216 models were trained, including the combinations for 6 $\times$ 4 $\times$ 3 choices for dimension reduction and 3 candidate mapping models. The prediction accuracy among these models were compared to determine the optimal one. Mean absolute error (MAE) was used as the evaluation metrics to identify the optimal model. Since any one mapping model is not necessarily achieves the best performance on all the parameters to estimate, the evaluation was conducted separately for each parameter, resulting in three optimal estimation models that tailored to each individual cell parameter. As a comparison, the conventional curve fitting~\citep{jiang2021mr} was also used as a reference for cell parameters estimation. {\rev To further investigate the relationship between optimal PCA dimensionality and sequence sensitivity, we performed additional studies that used each imaging sequence individually along with dimension reduction to estimate each cell microenvironment parameter, and the corresponding estimation MAE was compared to assess the relative sensitivity of each sequence.}

\subsection{Evaluation studies}
\subsubsection{Uncertainty quantification}
We first quantified the accuracy of our uncertainty estimation derived via the Jacobian approach. As such, we conducted a fitting experiment to estimate the five microenvironment parameters from dMRI signals. The estimation uncertainties were then calculated and compared with the derived uncertainties. In this step, we considered five values for each microenvironment parameter in its corresponding range, and each with 10 noise realizations in computing the dMRI signals. For each generated noisy dMRI signal, we derived the five microenvironment parameters by fitting the model in Eq.~(\ref{Eq_S}) with curve fitting using nonlinear least-squares optimization. The initial parameter values were set to 13~$\mu$m for $d$, 1.5~$\mu$m$^2$/ms for $D_{in}$, 50\% for $V_{in}$, 1.5~$\mu$m$^2$/ms for $D_{ex}$, and 5~$\mu$m$^2$ for $\beta_{ex}$ for the fitting. We computed the error between the obtained and the ground truth parameter values. 

 Based on these results, for each microenvironment-parameter combination, the uncertainty of each estimated parameter was quantified as the standard deviation (STD) of the fitting error across the 10 noise realizations. The final STD was then defined as the median of these STDs over the range of microenvironment-parameter combinations and compared with the corresponding value obtained in the theoretical analysis.


To further obtain insights about the uncertainty analysis results, we studied how variations in the five cell microenvironment parameters affect the dMRI signal. Generally, the more sensitive the dMRI signal to the microenvironment parameter is, the more robust it is to recover this microenvironment parameter. For this purpose, the dMRI signals were simulated using the analytical expression in Eq. \eqref{Eq_S}. We randomly varied one parameter of interest within the constrained range at a time, while fixing the other four parameters. The resulting signal variations were recorded and visualized. 


\subsubsection{Microenvironment parameters estimation model}

For the development of microenvironment parameter estimation models, we tested the 216 candidate models for each of the parameters and selected the best-performed ones as our final models. For these models, we conducted simulations to assess the accuracy and robustness of the estimation results. Specifically, an independent dataset was generated using the signal model in Eq.~(\ref{Eq_S}). These data were computed for 500 microenvironment parameters randomly selected in the interested parameter range. For each data, random noise was added 10 times, resulting in 5000 data pairs. Each noisy dMRI signal was then input into the well-trained optimal methods for prediction and the prediction results were evaluated by comparing with the known ground truth parameter value used in data generation. The model performance was further compared with that of the conventional method by fitting the dMRI signal, and the other two candidate methods that we considered but were found sub-optimal. Quantitatively, accuracy was evaluated by using the MAE and mean squared error (MSE). 

As for the evaluation on model robustness, we characterized this using the STD  of the 500 parameters set calculated over the 10 random noise realizations for a given ground truth microenvironment parameter,  leading to 500 STD samples. We further computed the mean values of the STDs over the  500 STDs as a metric to measure the overall model robustness. To compare model robustness among different models,  a two-sided Wilcoxon signed-rank test was performed, with $p<0.05$ considered statistically significant. To further assess the effect of noise, an additional study was conducted for a higher noise level of $\nu=0.05$,  $\nu=0.066$, and $\nu=0.10$, corresponding to a clinically achievable SNR of 20,  15, and 10. 

\subsubsection{\textit{In vitro} evaluation}

Finally, to evaluate the performance of the developed parameter estimation models, we conducted an \textit{in vitro} experiment using the well-established colorectal cancer cell line (MC38). The MC38 line is a widely used preclinical model for studying colorectal cancer and immunotherapies \citep{hos2020identification}. In this study, MC38 cells were cultured in complete DMEM (supplemented with 10\% FC II and 1x Penicillin-Streptomycin). Cells were allowed to grow until reaching confluency, at which point they were trypsinized, collected, centrifuged at 500 g, and counted. 30 million cells were collected and resuspended in 0.5 mL of media in 1.5 mL Eppendorf tube. The cell suspension was allowed to settle naturally at the bottom of the tube before imaging. The tube was scanned on a clinical 1.5~T MRI system (Ingenia Ambition X, Philips Healthcare, Netherlands). The dMRI signal was acquired using the PGSE, OGSEn1 and OGSEn2 sequences with parameters listed in Table~\ref{tab:pameters}. Other relevant scan parameters included TR 4500 ms, TE 164 ms, FOV 192 $\times$ 192 mm$^2$, voxel size 1.5 $\times$ 1.5 $\times$ 5 mm$^3$, and  number of slices: 5. A SENSE factor of 3 and half scan factor 0.63 were used. The total scan time was 5 min 42 s for PGSE, 3 min 54 s for OGSEn1, and 3 min for OGSEn2. The acquired dMRI signals were fed to the derived microenvironment parameter estimation model to derive the parameters of interest. For comparison, the parameters were also estimated using other models considered in this study. The estimated cell sizes $d$ were compared with the experimentally measured ground-truth values, which was derived by counting the cells using the TC20\textsuperscript{TM} automated cell counter (Bio-Rad Laboratories, Inc. Hercules, CA). However, other ground truth parameters were not available in the experiments due to lack of reliable methods to measure them.

\section{Results}
\label{sec:results}

\subsection{Uncertainty quantification}
\label{Sec:derivable parameters identification}

Table~\ref{tab:derivable_parameter_identification} summarizes the absolute ($\sigma_i$) and relative ($\delta_i$) uncertainty of estimated microenvironment parameters. Based on the analysis, the relative uncertainty values for the five parameters were 0.98 for $d$, 2.9 for $D_{in}$, 0.21 for $V_{in}$, 0.34 for $D_{ex}$, and 7.52 for $\beta_{ex}$. The relatively high relative uncertainties, considering the threshold of $\delta_0=1.0$, for $D_{in}$ and $\beta_{ex}$ indicate that the dMRI signals are not sensitive to these parameters, making it difficult to estimate these parameters accurately and resulting in limited robustness in their derivation, which is consistent with the findings of Xu et al., who reported that their fitting results were insensitive to the choice of $D_{in}$ and $\beta_{ex}$ \citep{jiang2016quantification,xu2021mri}. In contrast, the relatively low uncertainties for $d$, $V_{in}$, and $D_{ex}$ suggest that these parameters are mathematically robust and can be reliably estimated from the measured signals. 

The result in this table also confirmed the theoretical uncertainty estimation via numerical studies. The numerically computed uncertainties closely matched the corresponding derived values via the Jacobian analysis, with normalized uncertainties of 0.62 for $d$, 5.17 for $D_{in}$, 0.27 for $V_{in}$, 0.67 for $D_{ex}$, and 11.00 for $\beta_{ex}$. Overall, the numerically evaluated and analytically derived relative uncertainties generally agreed with each other, demonstrating the accuracy of our parameter uncertainty derivation.

\vspace{10pt}
\begin{table}[htbp]
    \raggedright
    \captionv{10}{}{Derived and numerically evaluated absolute ($\sigma_i$) and relative ($\delta_i$) uncertainty of estimated microenvironment parameters. Numbers in bold face indicate those parameters considered as robust.}
    \label{tab:derivable_parameter_identification}
    \renewcommand{\arraystretch}{1.5}
    \resizebox{1\linewidth}{!}{
        \renewcommand{\arraystretch}{}
        \begin{tabular}{c|c|ccccc}
            \hline\hline
            & 
            & $d$ ($\mu$m) 
            & $D_{in}$ ($\mu$m$^2$/ms)
            & $V_{in}$
            & $D_{ex}$ ($\mu$m$^2$/ms)
            & $\beta_{ex}$ ($\mu$m$^2$) \\
            \hline
            \multirow{2}{*}{\makebox[0.8cm][c]{$\sigma_i$}}
            & Derived
            & 12.78 & 5.19 & 10.43\% & 0.62 & 37.62 \\
            &Fitting
            & 8.08 & 9.25 & 13.32\% & 1.20 & 54.99 \\
            
            \hline\hline
            & 
            & $d$ 
            & $D_{in}$
            & $V_{in}$
            & $D_{ex}$
            & $\beta_{ex}$ \\
            \hline
            \multirow{2}{*}{\makebox[0.8cm][c]{$\delta_i$}} & Derived
            & \textbf{0.98} & 2.90 & \textbf{0.21} & \textbf{0.34} & 7.52 \\
            &Fitting
            & \textbf{0.62} & 5.17 & \textbf{0.27} & \textbf{0.67} & 11.00 \\
            \hline\hline
        \end{tabular}
    }
\end{table}

\vspace{15pt}
\begin{figure}[hbtp]
   \begin{center}
   \includegraphics[width=16.5cm]{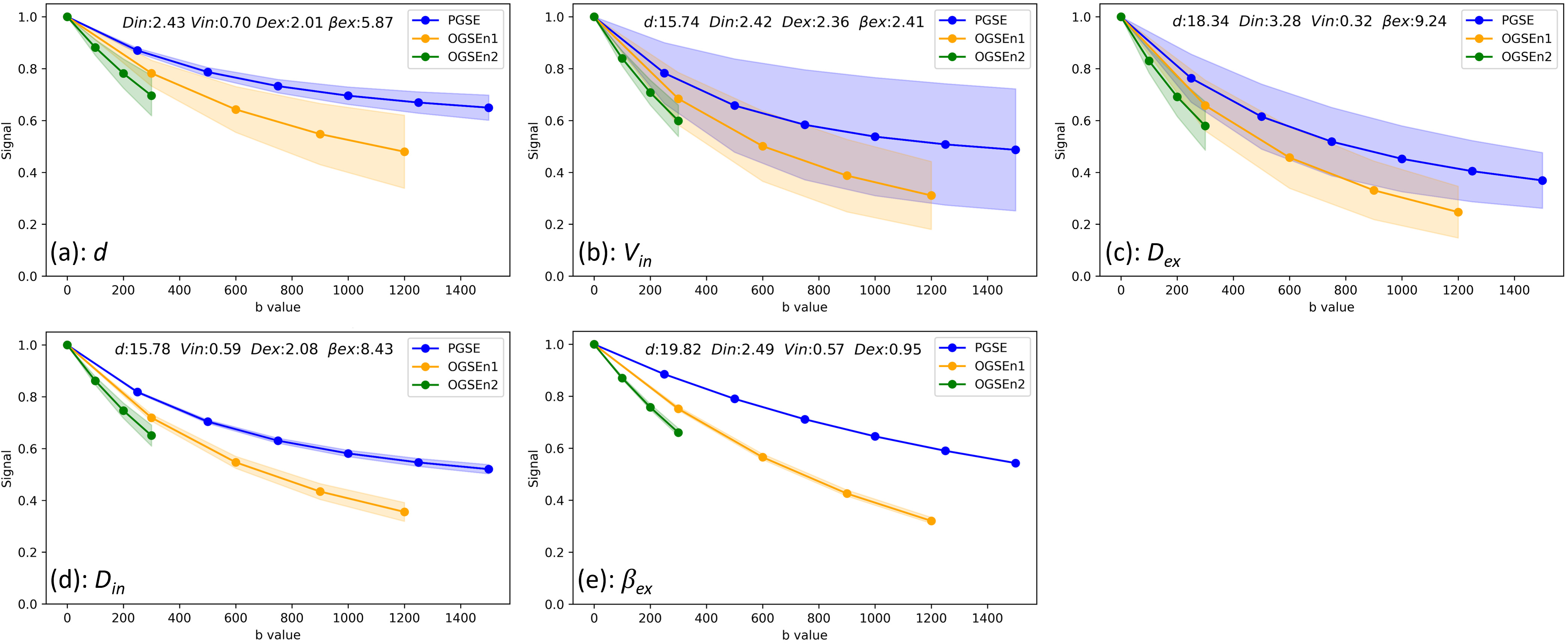}
   \captionv{14}{}{Illustration of signal variations with respect to microenvironment parameters. Panels (a)-(c) show the signal variations with respect to $d$, $V_{in}$, and $D_{ex}$, respectively, while keeping other parameters fixed at the values indicated in each subfigure. Panels (d) and (e) depict the signal variations with respect to $D_{in}$ and $\beta_{ex}$, respectively, with all other parameters fixed. The blue, orange, and green curves (and their corresponding shaded bands) represent the mean and standard deviation of the PGSE, OGSEn1, and OGSEn2 signals, respectively.}
    \label{parameter_validation} 
    \end{center}
\end{figure}

Figure~\ref{parameter_validation} illustrates the impact of each microenvironment parameter on the dMRI signals. As shown, relatively more pronounced signal variations were observed when changing $d$, $V_{in}$, and $D_{ex}$, whereas variations in $D_{in}$ and $\beta_{ex}$ produced only minimal changes. This behavior is consistent with our uncertainty analysis, which predicted that different microenvironment parameters contribute unequally to the overall signal variation. Specifically, parameters that exert a larger influence on the signal, such as $d$, $V_{in}$, and $D_{ex}$, introduce more distinguishable patterns in the measured data, making them relatively easier to infer through the mapping model. In contrast, parameters with weak signal sensitivity, such as $D_{in}$ and $\beta_{ex}$, have limited impact on the observed signal variations, which can make their accurate estimation more challenging.

\subsection{Microenvironment parameter estimation model}
\label{Sec:model development}

Based on the identified microenvironment parameters that can be estimated with the highest robustness, namely cell diameter $d$, intracellular volume fraction $V_{in}$, and extracellular diffusion coefficient $D_{ex}$, we developed dedicated parameter estimation models targeting these three parameters. 

\begin{figure}[hbtp!]
   \begin{center}
   \includegraphics[width=16.5cm]{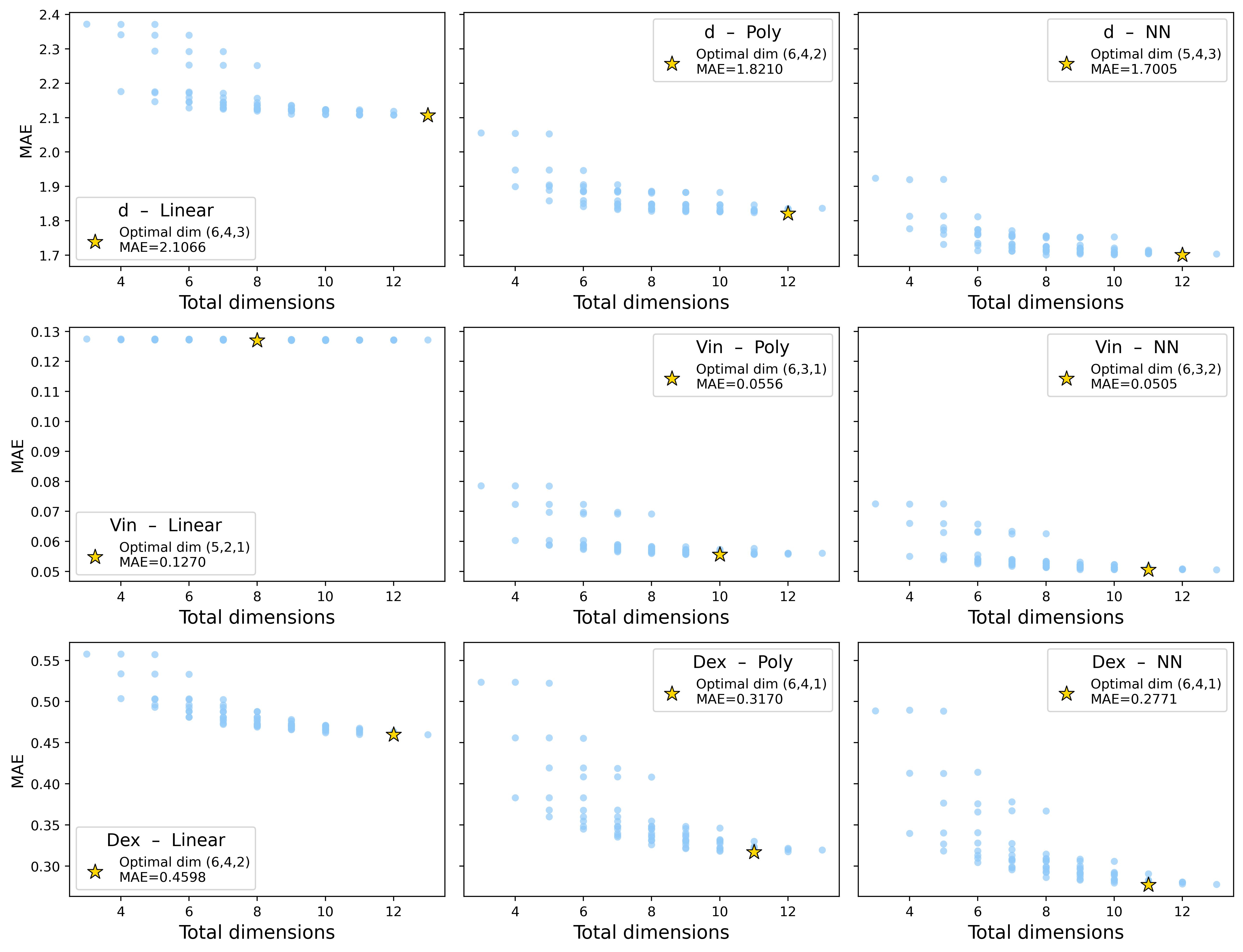}
   {\rev \captionv{14}{}{MAE as a function of total dimensions. Each panel corresponds to one target parameter ($d$, $V_{in}$, $D_{ex}$) and regression model (Linear: linear regression, Poly: polynomial regression, NN: 4-layer neural network). Each point represents one combination result. The star marks the selected optimal dimension result.}}
    \label{dimension_combinations} 
    \end{center}
\end{figure}

{\rev Figure~\ref{dimension_combinations} shows the MAE as a function of total dimensions summed over all the dimension over the three sequences. In each subfigure, there are 72 points corresponding to the MAEs under the 72 possible combinations of reduced dimensions for the three sequences. The identified optimal combination, shown as the star, appeared as the minimal among all the data points.} 

Among the 216 candidate models evaluated, the 4-layer neural network consistently demonstrated the best performance across all three parameters. This is likely due to the strong function approximation capability and flexibility of neural network architectures. The performance of the models on an independent test dataset is summarized in Table~\ref{comparison}. For each parameter and model, the triplet numbers correspond to the optimal number of PCA-reduced dimensions used for the PGSE, OGSEn1, and OGSEn2 sequences, respectively. For the 4-layer neural network, cell diameter $d$ estimation achieved optimal accuracy with reduced dimensions of $(5, 4, 3)$, yielding an MAE of 1.70 $\mu$m and an MSE of 2.85 $\mu$m$^2$. Estimation of $V_{in}$ performed best with reduced dimensions of $(6, 3, 2)$, resulting in an MAE of $5.05\%$ and an MSE of $0.76\%$. Finally, $D_{ex}$ estimation achieved minimal error with reduced dimensions of $(6, 4, 1)$, yielding an MAE of 0.28  $\mu$m$^2$/ms and an MSE of 0.18  $\mu$m$^4$/ms$^2$.

\vspace{10pt}
\begin{table}[htbp]
    \raggedright
    \captionv{10}{}{Comparison of parameter estimation model performances among different models. The triplet numbers in parentheses below each parameter indicate the optimal dimensional reductions for PGSE, OGSEn1, and OGSEn2, respectively. Numbers in bold face indicate the best performance.}
    \label{comparison}
    {\Large
    \resizebox{1\linewidth}{!}{
        \renewcommand{\arraystretch}{1.5}
        \begin{tabular}{lcccccccccccc}
            \hline\hline
            & \multicolumn{3}{c}{Fitting} 
            & \multicolumn{3}{c}{Linear Regression}
            & \multicolumn{3}{c}{Polynomial Regression} 
            & \multicolumn{3}{c}{4-layer Neural Network} \\
            \hline
            & $d (\mu\text{m})$ 
            & $V_{in}$ 
            & $D_{ex} (\mu\text{m}^2/\text{ms})$
            & $d (\mu\text{m})$ 
            & $V_{in}$ 
            & $D_{ex} (\mu\text{m}^2/\text{ms})$ 
            & $d (\mu\text{m})$ 
            & $V_{in}$ 
            & $D_{ex}  (\mu\text{m}^2/\text{ms})$ 
            & $d (\mu\text{m})$ 
            & $V_{in}$ 
            & $D_{ex} (\mu\text{m}^2/\text{ms})$ \\
            & - & - & -
            & (6-4-3) & (5-2-1) & (6-4-2)
            & (6-4-2) & (6-3-1) & (6-4-1)
            & (5-4-3) & (6-3-2) & (6-4-1) \\
            \hline
            MAE
            & 2.67 & 8.59\% & 0.52 
            & 2.11 & 12.71\% & 0.46 
            & 1.82 & 5.56\%  & 0.32 
            & \textbf{1.70} & \textbf{5.05\%}  & \textbf{0.28} \\
            MSE
            & 7.30 & 1.79\% & 0.62
            & 3.72 & 2.69\% & 0.36 
            & 3.08 & 0.85\% & 0.21 
            & \textbf{2.85} & \textbf{0.76\%} & \textbf{0.18} \\
            \hline\hline
        \end{tabular}
    }}
\end{table}


There are a few important observations in this table worth highlighting. First, the conventional fitting-based approach achieved better estimation accuracy for $V_{in}$ than linear regression, but it was outperformed by both polynomial regression and the 4-layer neural network. This can be attributed to the inherent limitation of linear models, which lack the flexibility to approximate the nonlinear relationships between the dMRI signals and the underlying microenvironment parameters. Second, overall, all three learning-based methods demonstrated superior performance compared to the conventional fitting approach. This improvement can be largely ascribed to the use of logarithmic signal transformation and dimension reduction, which enabled the models to focus on the most informative components of signal variation while suppressing noise effects.

\begin{figure}[hbtp!]
   \begin{center}
   \includegraphics[width=16.5cm]{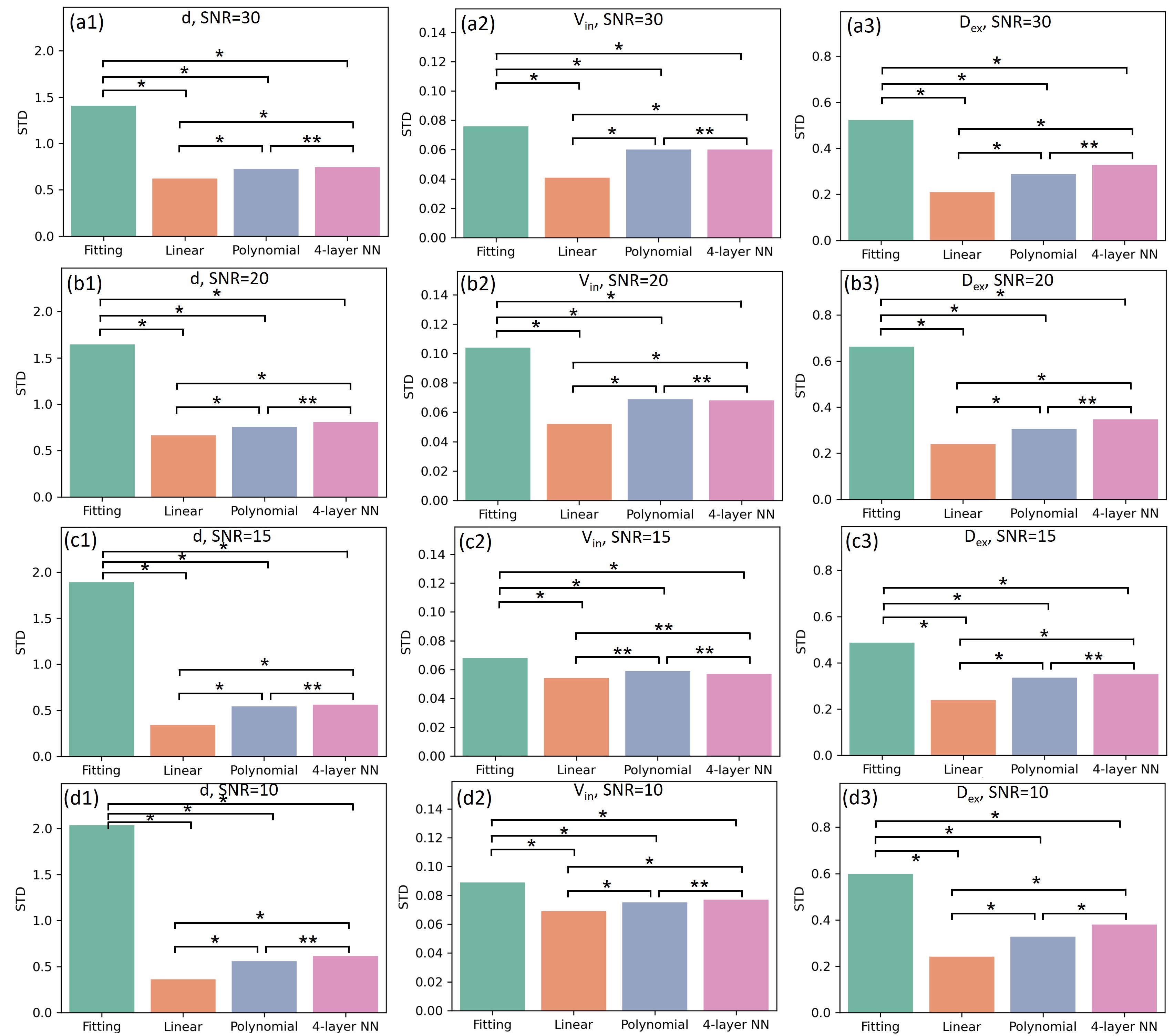}
   \captionv{14}{}{Microenvironment parameter estimation robustness comparison among the four models. Figures (a1)-(a3), (b1)-(b3), (c1)-(c3), and (d1)-(d3) show the results at SNR = 30, 20, 15, and 10, respectively. $*: p<0.05$; $**: p>0.05$.}
    \label{STD_comparison} 
    \end{center}
\end{figure}

Figure~\ref{STD_comparison} summarizes the robustness behaviors of different models and compares their performances. First, the conventional fitting method exhibited substantially larger STDs with statistical significance ($p<0.05$) than all the three learning-based methods for every parameter, indicating its inferior robustness. Among the learning-based approaches, the 4-layer neural network achieved average STDs of 0.747  $\mu$m (95\% CI: 0.719-0.774 $\mu$m) for $d$ estimation,  6.0\% (95\% CI: 5.7-6.3\%) for $V_{in}$ estimation, and 0.328  $\mu$m$^2$/ms (95\% CI: 0.318-0.347 $\mu$m$^2$/ms) for  $D_{ex}$ estimation. No statistically significant differences ($p > 0.05$) were observed when comparing these values with those from the polynomial regression models. Although linear regression yielded the lowest STDs with statistical significance for all three parameters, its relatively poor estimation accuracy, as discussed above, makes it less suitable for practical use. Taking both accuracy and robustness into account, our results indicated that the 4-layer neural network is the preferred model for estimating microenvironment parameters from dMRI signals.  The same conclusion was observed for the datasets with SNR = 20, 15, and 10, as shown in Figure~\ref{STD_comparison}(b-d). Notably, the estimation of $D_{ex}$ was more affected at SNR = 10, suggesting that this parameter is particularly sensitive to severe noise contamination and highlighting a limitation of the neural network under extremely low-SNR conditions.

{\rev Table~\ref{tab:model_comparison} summarizes the sequence-specific cell parameter sensitivity analysis results. According to our findings, OGSEn1 demonstrates higher sensitivity to cell diameter than PGSE and OGSEn2, even though its optimal PCA dimension is lower than that of PGSE. This observation is physically reasonable because OGSE sequences are generally more sensitive to smaller cells. In particular, OGSEn2 employs a higher oscillation frequency, resulting in sensitivity to even smaller structural scales~\citep{jiang2016quantification}. In contrast, PGSE showed higher sensitivity to $V_{in}$ and $D_{ex}$ than OGSEn1 and OGSEn2. We believe this is because $V_{in}$ and $D_{ex}$ primarily influence compartment-level diffusion rather than the structural length scales. PGSE with longer diffusion times accumulate stronger contrast between intracellular and extracellular environments, thereby improving the estimation accuracy of $V_{in}$ and $D_{ex}$. This observation was consistently observed across all three learning-based models.}

\begin{table}[h]
\centering
{\rev \caption{MAE results for evaluating the sensitivity of each diffusion sequence to different microenvironment parameters. Dim: optimal PCA dimension.}}
\label{tab:model_comparison}
\begin{tabular}{llccc}
\toprule
\textbf{Param} & \textbf{Model} & \textbf{PGSE (Dim)} & \textbf{OGSEn1 (Dim)} & \textbf{OGSEn2 (Dim)} \\
\midrule
\multirow{3}{*}{$d$ ($\mu$m)}
  & Linear       & 3.48 (6) & \textbf{3.15 (4)} & 3.33 (1) \\
  & Polynomial   & 3.26 (6) & \textbf{2.62 (4)} & 3.13 (2) \\
  & 4-layer NN   & 3.15 (5) & \textbf{2.50 (4)} & 3.05 (3) \\
\midrule
\multirow{3}{*}{$V_\text{in}$}
  & Linear       & \textbf{0.13 (5)} & 0.16 (4) & 0.20 (3) \\
  & Polynomial   & \textbf{0.08 (6)} & 0.15 (4) & 0.20 (1) \\
  & 4-layer NN   & \textbf{0.07 (5)} & 0.15 (3) & 0.19 (2) \\
\midrule
\multirow{3}{*}{$D_\text{ex}$ ($\mu$m$^2$/ms)}
  & Linear       & \textbf{0.50 (6)} & 0.55 (4) & 0.60 (2) \\
  & Polynomial   & \textbf{0.39 (6)} & 0.50 (4) & 0.59 (1) \\
  & 4-layer NN   & \textbf{0.34 (6)} & 0.45 (4) & 0.57 (1) \\
\bottomrule
\end{tabular}
\end{table}

\subsection{\textit{In vitro} experiment validation}
\label{Sec:in vitro experiment}

\begin{figure}[hbtp!]
   \begin{center}
   \includegraphics[width=16.5cm]{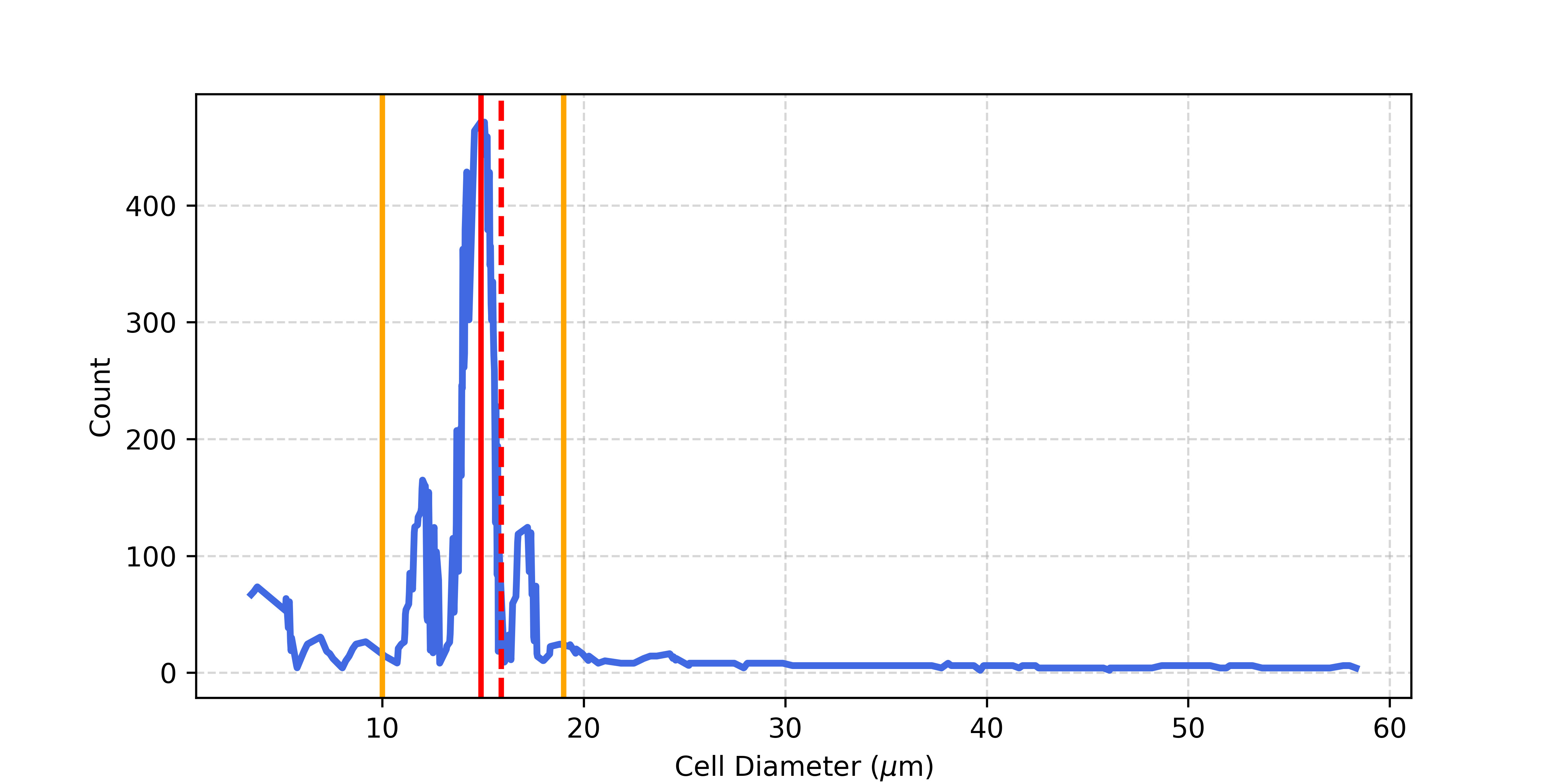}
   \captionv{14}{}{Cell diameter distribution of the MC38 cells. The two yellow lines indicate the selected cell size gates (10 $\mu$m - 19 $\mu$m), and the red line shows the average cell diameter (15.0 $\mu$m). Dash line indicate the cell diameter (16.0 $\mu$m) derived from the 4-layer neural network model.}
    \label{Cell_distribution} 
    \end{center}
\end{figure}

The measured cell diameter distribution using the automated cell counter is illustrated in Figure~\ref{Cell_distribution}, which shows an average cell diameter of approximately 15.0 $\mu$m. Using dMRI signals acquired from our 1.5T clinical scanner and applying the 4-layer neural network parameter estimation model, we achieved the cell diameter 16.0 $\mu$m, corresponding to a relative difference of 6.67\%, indicating that the 4-layer neural network is promising for cell size estimation in practice. The cell diameters estimated by fitting, linear regression, and polynomial regression were 13.6 $\mu$m, 13.0 $\mu$m, and 16.1 $\mu$m, respectively.

Additionally, we also estimated an intracellular volume fraction of 34.8\% and an extracellular diffusion coefficient of 1.0 $\mu$m$^2$/ms using the developed optimal models. Although ground truth experimental values for these two parameters are currently unavailable, they were considered as reasonable values. We roughly estimated $V_{in}$ based on the following experimental setup: 30 million cells were resuspended in 0.5 mL of media, and after removing the supernatant following centrifugation, 0.15 mL remained. Given a ground truth cell diameter of 15 $\mu$m, the intracellular volume fraction $V_{in}$ was thus estimated to be 35.4\%. Moreover, tumor microenvironment diffusion models (e.g. VERDICT and IMPULSED) reported extracellular diffusivities on the order of 1.0 $\mu$m$^2$/ms in dense cellular environments \citep{panagiotaki2014noninvasive,xu2020magnetic}. 

\section{Discussions}

\subsection{Clinical Significance}
In this study, we quantified the uncertainty of cell microenvironment parameter estimation and proposed a framework for robust and accurate parameter estimation. The intended application is radiotherapy response assessment, which has significant potential for clinical translation. {\rev Specifically, the radiation induced apoptosis, the dominant model of cell death at low-to-moderate doses, is characterized by cell shrinkage and nuclear condensation, which would manifest as a reduction in $d$ and a decrease in $V_{in}$~\citep{klassen1993two,jiang2019vivo}. At higher dose or in radioresistant tumor regions, necrosis predominates, which involves cell swelling, membrane disruption, and subsequent release of intracellular contents, these changes are expected to increase extracellular water mobility and thus increase $D_{ex}$~\citep{schlemmer2002differentiation,lee2018regulation}. Furthermore, radiation-induced immune cell infiltration introduces competing microstructural effects, including increased local cellularity and reduced extracellular space, that may be disentangled from tumor cell death by the concomitant decrease in $d$~\citep{zhu2022radiotherapy}. Future validation experiments may include longitudinal diffusion MRI measurements during radiotherapy combined with histopathological analyses, such as H$\&$E staining, apoptosis assays (e.g., TUNEL), necrosis quantification, and immune cell marker staining, to establish correlations between the estimated diffusion-derived parameters and underlying biological responses.} By enabling the non-invasive estimation of key cellular microenvironment parameters, it offers an alternative to conventional biopsy procedures, which are often invasive, costly, and difficult to repeat. This capability is particularly valuable in clinical scenarios where repeated biopsies for treatment monitoring are impractical or impossible, such as in deep-seated tumors, anatomically sensitive locations, or medically fragile patients.  Furthermore, integrating these imaging-derived biomarkers into radiotherapy workflows could enable longitudinal monitoring of tumor microenvironment before, during and after treatment, providing early indicators of therapeutic response and potentially guiding adaptive treatment strategies.\citep{matuszak2019functional,nguyen2022advances,li2013automatic}


This study provides important theoretical insight into which cellular microenvironment parameters can be reliably estimated from dMRI signals. Through uncertainty analysis, we demonstrated that cell diameter $d$, intracellular volume fraction $V_{in}$, and extracellular diffusion coefficient $D_{ex}$ can be robustly derived from measured dMRI data, whereas the other two parameters, $D_{in}$ and $\beta_{ex}$, are more challenging to estimate accurately. This difference arises from the distinct mathematical roles these parameters play in the signal model, which determine their relative impact on the measured signal and, consequently, the achievable estimation precision. Among these parameters that can be robustly estimated, $d$ and $V_{in}$ are closely linked to the morphological properties of the cellular microenvironment. Radiotherapy is known to induce morphological changes at the microscopic level, such as cell shrinkage during apoptosis or alterations in cell density due to immune cell infiltration. Therefore, the demonstrated robustness of $d$ and $V_{in}$ supports their potential use as reliable quantitative measures of microstructural alterations in response to therapy.

Importantly, the theoretical guidance regarding which parameters can be estimated with confidence  provides a foundation for future research to optimize imaging protocols and develop targeted estimation algorithms. Notably, our findings are consistent with previous empirical observations \citep{jiang2016quantification,xu2021mri}, thereby offering a strong theoretical basis for parameter selection in microenvironmental imaging studies and paving the way for non-invasive, quantitative monitoring of tumor response at the microscopic scale.

\subsection{Methodological Considerations}
Building upon the parameters identified as robustly estimable through analysis, we further developed direct mapping models to estimate cellular microenvironment parameters from dMRI signals, integrating logarithmic signal transformation, dimensionality reduction, and a neural network estimator to learn the relationship from measured signals to underlying parameters. Compared with the conventional fitting-based approach, the mapping model offers several important advantages. First, by transforming and reducing the input data, the model focuses on the most informative variations in the signal while suppressing noise, leading to improved estimation stability. Second, the direct mapping circumvents the ill-posed nature and noise amplification problems often encountered in nonlinear optimization during signal fitting. Third, once trained, the mapping model enables fast inference, allowing parameter estimation to be performed efficiently without iterative fitting procedures. Collectively, this strategy provides a more robust, accurate, and computationally efficient alternative to conventional methods, with clear potential for integration into clinical imaging workflows. 

 Dimensionality reduction via PCA is a key step in our model to balance information retention against noise suppression. Rather than selecting an optimal dimension in isolation, we coupled dimensionality selection with the subsequent neural network to minimize the MAE of the entire prediction pipeline. This two-stage design allows PCA serves as a preprocessing step to reduce noise and redundant information, while the downstream model performs the mapping from the compressed signal representation to the target microenvironment parameters.

 We employed a Rician noise model to realistically reflect the signal-dependent nature of the noise signal in the magnitude MR images. While Rician noise can introduce bias at low SNR, potentially affecting linear methods like PCA, prior research indicates it is well-approximated by a Gaussian distribution when the SNR exceeds $\sim$5 \citep{gudbjartsson1995rician}. Since our study operates within this higher SNR range, the Rician-induced bias is negligible. Consequently, the combination of logarithmic transformations and PCA remains an effective approach for handling the noise characteristics in our simulations.

 In recent years, deep learning has transformed numerous scientific and clinical domains through its ability to handle complex data and capture intricate nonlinear relationships. Its effectiveness has been demonstrated in a wide range of applications, including disease diagnosis \citep{sharma2022deep}, medical imaging \citep{yinghui2025artificial,li2024evaluating}, and parameter estimation tasks \citep{karimi2021deep}. However, excessive architectural complexity can increase the risk of overfitting and estimation instability. In this study, we chose a 4-layer fully connected network as the balance between capacity and robustness. Since our model input consists of low-dimensional feature vectors derived from PCA, rather than raw image data with spatial dependencies, the specific advantages of more complex architectures like convolutional neural networks or transformers were not required. Under this setting, a 4-layer fully connected network provided sufficient model capacity to capture the nonlinear mapping while maintaining relatively low model complexity and reducing risk of overfitting, as preliminary experiments with more complex networks yielded no measurable performance gains.

\subsection{Study Limitations and Future Directions}
 \begin{itemize}
 \item In our current study, each parameter was estimated using a separated model. Yet the uncertainty analysis revealed covariances between parameters. We have previously addressed a similar issue in the context of {\rev Intravoxel Incoherent Motion (IVIM)} parameter estimation \citep{dai2025r}, where we introduced a composite metric (R-index) that accounts for covariance among various estimated parameters. We are currently extending this approach to the present model to address this issue.
 
 \item {\rev The sequence-specific cell parameter sensitivity analysis experiment provides insight into the relative contribution of each diffusion sequence to parameter estimation. We observed that the MAE obtained using a single diffusion sequence was consistently higher than that achieved using all three sequences combined. This highlights the complementary information provided by different diffusion sequences and demonstrates the benefit of multi-sequence for improving parameter estimation accuracy. This work only investigated the three sequences that have used in previous studies \citep{xu2014mapping,jiang2016quantification}. It is possible that additional sequences could provide more valuable information to further enhance parameter estimation performance. Identifying the optimal combination of diffusion sequences and acquisition settings remains an important topic for future research and is beyond the scope of the current study. }

 \item The current protocol's acquisition time of approximately 12 minutes poses a significant barrier to routine clinical adoption. To address this, we recently developed an protocol optimization framework using {\rev Expected Information Gain (EIG)} and Bayesian optimization to maximize the information content across the high-dimensional acquisition space, including pulse types (PGSE and OGSE), diffusion times, and $b$-values \citep{dai2026optimizingimpulsedacquisitionprotocols}. This strategy demonstrated that scan times can be reduced to approximately 4 minutes without compromising estimation performance in simulation studies. {\rev Although the optimization framework was demonstrated using scanner settings representative of a 3T system, the framework itself can be adapted to 1.5T scanners by updating scanner-dependent constraints, particularly maximum achievable amplitude of diffusion encoding gradient and measurement SNR, followed by re-optimization under the corresponding acquisition conditions.} Ongoing work is focused on further confirming the feasibility and robustness of this optimized protocol in clinical practice.

 \item Our study focused on a specific dMRI imaging protocol and SNR setting under a 1.5T scanner setting, because the 1.5 T scanners are often used in radiation oncology applications. This setting generally leads to more challenging for diffusion MRI due to the lower SNR than the more commonly available 3.0T scanners in radiology applications. Therefore, demonstrating the feasibility of the proposed pipeline under 1.5 T conditions provides an initial validation in a relatively demanding scenario. In principle, the same pipeline can be extended to 3.0T scanners or modified protocols. However, the model’s performance under different field strengths or sequence settings should be further validated. Although 3.0T generally provides higher SNRs, it may also introduce other sources of variation like differences in tissue contrast, geometric distortion, B0/B1 inhomogeneity, and sequence-dependent signal characteristics. These factors may affect the performance of a model trained on 1.5 T data. To improve generalizability, future work will incorporate domain adaptation strategies, including retraining under different acquisition conditions, transfer learning, and normalization of signal distributions across scanners and protocols.

 \end{itemize}

\subsection{Experimental Validation}
When applying our derived parameter estimation model built using synthesized data to our \textit{in vitro} validation experiment, we observed an estimation error of approximately 1~$\mu$m for the cell diameters. This difference may arise from multiple factors. First real MRI scans are subject to complex factors such as magnetic field inhomogeneities, gradient non-linearities, which are difficult to fully replicate and hence were ignored in our simulations. The parameter estimation model built from simulation data was hence not able to be fully applicable to the real data. Additionally, our model assumed a single cell size, which was in contrast to the experimental setting with cell size distributed in a range. Nevertheless, this level of discrepancy is considered acceptable. Xu et al. \citep{xu2021mri} reported a 1.6~$\mu$m difference between real and estimated cell diameters in their experiments, which is of similar magnitude.

 Due to experimental constraints, this study focused exclusively on validating the cell size estimation. Future work should include more comprehensive validations on $V_{in}$ and $D_{ex}$. {\rev In our study, the validation of $V_{in}$ was performed indirectly using cell counting and volume estimation rather than direct measurement. Therefore, the estimated $V_{in}$ values may be affected by several sources of uncertainty. For example, cell sedimentation during sample preparation may lead to nonuniform cell distribution and inaccurate estimation of cell density and intracellular water fraction. Cell lysis or membrane damage may alter the effective intracellular volume and introduce discrepancies between the estimated and actual $V_{in}$ values. Additionally, the volume calculation assumes ideal spherical cell geometry, whereas real cells may exhibit irregular shapes and size heterogeneity. Such deviations from the spherical assumption could introduce bias into the estimated intracellular water fraction. In the future study, } $V_{in}$ may be evaluated using high-resolution microscopy, for example by estimating the cellular area fraction from microscopic images for comparison with the MRI-derived $V_{in}$. For $D_{ex}$, direct measurement is more challenging. It may be assessed indirectly using well-controlled phantoms or cell systems with known extracellular environments. In addition, the current MC38 suspension assumes relatively uniform packing and does not capture the heterogeneity that may exist in realistic biological tissues, which may alter the local intracellular volume fraction and extracellular diffusion environment, thereby changing the diffusion signal characteristics. Future studies with more heterogeneous and realistic sample configurations will therefore be important for further validation.

Moreover, the present method considers only a single cell type, whereas the tumor microenvironment is a heterogeneous system composed of multiple cell populations with distinct microenvironment properties, including tumor cells, immune cells, and endothelial cells.  Therefore, the current study should be considered a foundational methodological study for microenvironment parameter estimation. In future work, developing and validating multi-cell models will be an important direction to more accurately capture the complexity of realistic clinical scenarios.  To this end, animal studies may serve as useful investigations to further evaluate the proposed method under more complex microenvironment conditions that more closely reflect those in human subjects.

\section{Conclusions}

In conclusion, this study systematically investigated the uncertainty characteristics of cell microenvironment parameter estimation from dMRI signals and established a theoretical foundation for identifying which parameters can be robustly derived. Through analysis, we demonstrated that cell diameter $d$, intracellular volume fraction $V_{in}$, and extracellular diffusion coefficient $D_{ex}$ can be reliably estimated under clinically relevant imaging conditions, while intracellular diffusion coefficient $D_{in}$ and exchange rate $\beta_{ex}$ remain challenging due to the weak signal sensitivity with respect to them. Building on this finding, we developed a direct mapping framework that integrates logarithmic signal transformation, dimensionality reduction, and neural network modeling to estimate these three key parameters from dMRI signals. This approach demonstrated improved accuracy and robustness compared to conventional fitting methods, and its feasibility was further supported by \textit{in vitro} experimental validation. This work provides theoretical guidance for parameter selection, and potentially offers a practical and efficient solution for non-invasive, quantitative monitoring of tumor microenvironment during radiotherapy. 

\section*{Author Contribution Statement}
Conceptualization: Xun Jia and Jie Deng; Methodology: Xun Jia, Wen Li, Jie Deng, and Yan Dai; Software: Wen Li; Validation: Xun Jia and Wen Li; Formal analysis: Xun Jia, Jie Deng, and Wen Li; Investigation: Xun Jia and Wen Li; Resources: Xun Jia, Arely Perez Rodriguez, Todd Aguilera, and Jie Deng; Data curation: Arely Perez Rodriguez, Todd Aguilera, and Yan Dai; Writing-original draft preparation: Wen Li; Writing-review and editing: all authors; Supervision: Xun Jia and Jie Deng; Project administration: Xun Jia; Funding acquisition: Xun Jia. All authors have read and agreed to the published version of the manuscript.

\section*{Acknowledgement}
This study was supported in part by grants from NIH (R01EB032716, R37CA214639, R01CA227289, R01ACA285379).

\section*{Conflict of Interest Statement}
The authors have no relevant conflicts of interest to disclose.

\section*{References}






\vspace{-1cm}
\begin{thebibliography}{10}

\bibitem{therasse2000new}
P.~Therasse et~al.,
\newblock New guidelines to evaluate the response to treatment in solid tumors,
\newblock Journal of the national cancer institute {\bf 92}, 205--216 (2000).

\bibitem{newton2011progress}
R.~C. Newton, S.~V. Kemp, P.~L. Shah, D.~Elson, A.~Darzi, K.~Shibuya, S.~Mulgrew, and G.-Z. Yang,
\newblock Progress toward optical biopsy: bringing the microscope to the patient,
\newblock Lung {\bf 189}, 111--119 (2011).

\bibitem{swartz2018influence}
J.~E. Swartz, J.~P. Driessen, P.~M. van Kempen, R.~de~Bree, L.~M. Janssen, F.~A. Pameijer, C.~H. Terhaard, M.~E. Philippens, and S.~Willems,
\newblock Influence of tumor and microenvironment characteristics on diffusion-weighted imaging in oropharyngeal carcinoma: A pilot study,
\newblock Oral oncology {\bf 77}, 9--15 (2018).

\bibitem{bai2021study}
Y.~Bai et~al.,
\newblock Study of diffusion weighted imaging derived diffusion parameters as biomarkers for the microenvironment in gliomas,
\newblock Frontiers in Oncology {\bf 11}, 672265 (2021).

\bibitem{jordan2005dynamic}
B.~F. Jordan, M.~Runquist, N.~Raghunand, A.~Baker, R.~Williams, L.~Kirkpatrick, G.~Powis, and R.~J. Gillies,
\newblock Dynamic contrast-enhanced and diffusion MRI show rapid and dramatic changes in tumor microenvironment in response to inhibition of HIF-1$\alpha$ using PX-478,
\newblock Neoplasia {\bf 7}, 475--485 (2005).

\bibitem{xu2021mri}
J.~Xu et~al.,
\newblock MRI-cytometry: mapping nonparametric cell size distributions using diffusion MRI,
\newblock Magnetic resonance in medicine {\bf 85}, 748--761 (2021).

\bibitem{sigmund2011intravoxel}
E.~E. Sigmund, G.~Y. Cho, S.~Kim, M.~Finn, M.~Moccaldi, J.~H. Jensen, D.~K. Sodickson, J.~D. Goldberg, S.~Formenti, and L.~Moy,
\newblock Intravoxel incoherent motion imaging of tumor microenvironment in locally advanced breast cancer,
\newblock Magnetic resonance in medicine {\bf 65}, 1437--1447 (2011).

\bibitem{chenevert1997monitoring}
T.~L. Chenevert, P.~E. McKeever, and B.~D. Ross,
\newblock Monitoring early response of experimental brain tumors to therapy using diffusion magnetic resonance imaging.,
\newblock Clinical cancer research: an official journal of the American Association for Cancer Research {\bf 3}, 1457--1466 (1997).

\bibitem{xu2009quantitative}
J.~Xu, M.~D. Does, and J.~C. Gore,
\newblock Quantitative characterization of tissue microstructure with temporal diffusion spectroscopy,
\newblock Journal of magnetic resonance {\bf 200}, 189--197 (2009).

\bibitem{jiang2016quantification}
X.~Jiang, H.~Li, J.~Xie, P.~Zhao, J.~C. Gore, and J.~Xu,
\newblock Quantification of cell size using temporal diffusion spectroscopy,
\newblock Magnetic resonance in medicine {\bf 75}, 1076--1085 (2016).

\bibitem{stejskal1965use}
E.~O. Stejskal,
\newblock Use of spin echoes in a pulsed magnetic-field gradient to study anisotropic, restricted diffusion and flow,
\newblock The Journal of Chemical Physics {\bf 43}, 3597--3603 (1965).

\bibitem{callaghan1993principles}
P.~T. Callaghan,
\newblock {\em Principles of nuclear magnetic resonance microscopy},
\newblock Clarendon press, 1993.

\bibitem{cory1990measurement}
D.~Cory,
\newblock Measurement of translational displacement probabilities by NMR: an indicator of compartmentation,
\newblock Magnetic resonance in medicine {\bf 14}, 435--444 (1990).

\bibitem{brownstein1979importance}
K.~R. Brownstein and C.~Tarr,
\newblock Importance of classical diffusion in NMR studies of water in biological cells,
\newblock Physical review A {\bf 19}, 2446 (1979).

\bibitem{tanner1968restricted}
J.~E. Tanner and E.~O. Stejskal,
\newblock Restricted self-diffusion of protons in colloidal systems by the pulsed-gradient, spin-echo method,
\newblock The Journal of Chemical Physics {\bf 49}, 1768--1777 (1968).

\bibitem{yablonskiy2003statistical}
D.~A. Yablonskiy, G.~L. Bretthorst, and J.~J. Ackerman,
\newblock Statistical model for diffusion attenuated MR signal,
\newblock Magnetic Resonance in Medicine: An Official Journal of the International Society for Magnetic Resonance in Medicine {\bf 50}, 664--669 (2003).

\bibitem{zhao2008intracellular}
L.~Zhao, A.~Sukstanskii, C.~Kroenke, J.~Song, D.~Piwnica-Worms, J.~Ackerman, and J.~J. Neil,
\newblock Intracellular water specific MR of microbead-adherent cells: HeLa cell intracellular water diffusion,
\newblock Magnetic Resonance in Medicine: An Official Journal of the International Society for Magnetic Resonance in Medicine {\bf 59}, 79--84 (2008).

\bibitem{iima2014characterization}
M.~Iima, O.~Reynaud, T.~Tsurugizawa, L.~Ciobanu, J.-R. Li, F.~Geffroy, B.~Djemai, M.~Umehana, and D.~Le~Bihan,
\newblock Characterization of glioma microcirculation and tissue features using intravoxel incoherent motion magnetic resonance imaging in a rat brain model,
\newblock Investigative radiology {\bf 49}, 485--490 (2014).

\bibitem{ichikawa2018histological}
S.~Ichikawa, U.~Motosugi, D.~Hernando, H.~Morisaka, N.~Enomoto, M.~Matsuda, and H.~Onishi,
\newblock Histological grading of hepatocellular carcinomas with intravoxel incoherent motion diffusion-weighted imaging: inconsistent results depending on the fitting method,
\newblock Magnetic Resonance in Medical Sciences {\bf 17}, 168--173 (2018).

\bibitem{ades2018evaluation}
B.~Ades-Aron, J.~Veraart, P.~Kochunov, S.~McGuire, P.~Sherman, E.~Kellner, D.~S. Novikov, and E.~Fieremans,
\newblock Evaluation of the accuracy and precision of the diffusion parameter EStImation with Gibbs and NoisE removal pipeline,
\newblock Neuroimage {\bf 183}, 532--543 (2018).

\bibitem{oeschger2023axisymmetric}
J.~M. Oeschger, K.~Tabelow, and S.~Mohammadi,
\newblock Axisymmetric diffusion kurtosis imaging with Rician bias correction: A simulation study,
\newblock Magnetic Resonance in Medicine {\bf 89}, 787--799 (2023).

\bibitem{iima2021perfusion}
M.~Iima,
\newblock Perfusion-driven intravoxel incoherent motion (IVIM) MRI in oncology: applications, challenges, and future trends,
\newblock Magnetic Resonance in Medical Sciences {\bf 20}, 125--138 (2021).

\bibitem{xu2020magnetic}
J.~Xu et~al.,
\newblock Magnetic resonance imaging of mean cell size in human breast tumors,
\newblock Magnetic resonance in medicine {\bf 83}, 2002--2014 (2020).

\bibitem{shen2020introduction}
C.~Shen, D.~Nguyen, Z.~Zhou, S.~B. Jiang, B.~Dong, and X.~Jia,
\newblock An introduction to deep learning in medical physics: advantages, potential, and challenges,
\newblock Physics in Medicine \& Biology {\bf 65}, 05TR01 (2020).

\bibitem{li2022multi}
W.~Li et~al.,
\newblock Multi-institutional investigation of model generalizability for virtual contrast-enhanced MRI synthesis,
\newblock in {\em International Conference on Medical Image Computing and Computer-Assisted Intervention}, pages 765--773, Springer, 2022.

\bibitem{assaf2004new}
Y.~Assaf, R.~Z. Freidlin, G.~K. Rohde, and P.~J. Basser,
\newblock New modeling and experimental framework to characterize hindered and restricted water diffusion in brain white matter,
\newblock Magnetic Resonance in Medicine: An Official Journal of the International Society for Magnetic Resonance in Medicine {\bf 52}, 965--978 (2004).

\bibitem{alexander2010orientationally}
D.~C. Alexander, P.~L. Hubbard, M.~G. Hall, E.~A. Moore, M.~Ptito, G.~J. Parker, and T.~B. Dyrby,
\newblock Orientationally invariant indices of axon diameter and density from diffusion MRI,
\newblock Neuroimage {\bf 52}, 1374--1389 (2010).

\bibitem{xu2014mapping}
J.~Xu, H.~Li, K.~D. Harkins, X.~Jiang, J.~Xie, H.~Kang, M.~D. Does, and J.~C. Gore,
\newblock Mapping mean axon diameter and axonal volume fraction by MRI using temporal diffusion spectroscopy,
\newblock Neuroimage {\bf 103}, 10--19 (2014).

\bibitem{li2014fast}
H.~Li, J.~C. Gore, and J.~Xu,
\newblock Fast and robust measurement of microstructural dimensions using temporal diffusion spectroscopy,
\newblock Journal of magnetic resonance {\bf 242}, 4--9 (2014).

\bibitem{jiang2021mr}
X.~Jiang, H.~Li, S.~P. Devan, J.~C. Gore, and J.~Xu,
\newblock MR cell size imaging with temporal diffusion spectroscopy,
\newblock Magnetic resonance imaging {\bf 77}, 109--123 (2021).

\bibitem{hos2020identification}
B.~J. Hos et~al.,
\newblock Identification of a neo-epitope dominating endogenous CD8 T cell responses to MC-38 colorectal cancer,
\newblock Oncoimmunology {\bf 9}, 1673125 (2020).

\bibitem{panagiotaki2014noninvasive}
E.~Panagiotaki, S.~Walker-Samuel, B.~Siow, S.~P. Johnson, V.~Rajkumar, R.~B. Pedley, M.~F. Lythgoe, and D.~C. Alexander,
\newblock Noninvasive quantification of solid tumor microstructure using VERDICT MRI,
\newblock Cancer research {\bf 74}, 1902--1912 (2014).

\bibitem{klassen1993two}
N.~Klassen, P.~Walker, C.~Ross, J.~Cygler, and B.~Lach,
\newblock Two-stage cell shrinkage and the OER for radiation-induced apoptosis of rat thymocytes,
\newblock International journal of radiation biology {\bf 64}, 571--581 (1993).

\bibitem{jiang2019vivo}
X.~Jiang, E.~T. McKinley, J.~Xie, H.~Li, J.~Xu, and J.~C. Gore,
\newblock In vivo magnetic resonance imaging of treatment-induced apoptosis,
\newblock Scientific reports {\bf 9}, 9540 (2019).

\bibitem{schlemmer2002differentiation}
H.-P. Schlemmer, P.~Bachert, M.~Henze, R.~Buslei, K.~Herfarth, J.~Debus, and G.~Van~Kaick,
\newblock Differentiation of radiation necrosis from tumor progression using proton magnetic resonance spectroscopy,
\newblock Neuroradiology {\bf 44}, 216--222 (2002).

\bibitem{lee2018regulation}
S.~Y. Lee, M.~K. Ju, H.~M. Jeon, E.~K. Jeong, Y.~J. Lee, C.~H. Kim, H.~G. Park, S.~I. Han, and H.~S. Kang,
\newblock Regulation of tumor progression by programmed necrosis,
\newblock Oxidative medicine and cellular longevity {\bf 2018}, 3537471 (2018).

\bibitem{zhu2022radiotherapy}
S.~Zhu, Y.~Wang, J.~Tang, and M.~Cao,
\newblock Radiotherapy induced immunogenic cell death by remodeling tumor immune microenvironment,
\newblock Frontiers in immunology {\bf 13}, 1074477 (2022).

\bibitem{matuszak2019functional}
M.~M. Matuszak, R.~Kashani, M.~Green, C.~Lee, Y.~Cao, D.~Owen, S.~Jolly, and M.~Mierzwa,
\newblock Functional adaptation in radiation therapy,
\newblock in {\em Seminars in radiation oncology}, volume~29, pages 236--244, Elsevier, 2019.

\bibitem{nguyen2022advances}
D.~Nguyen, M.-H. Lin, D.~Sher, W.~Lu, X.~Jia, and S.~Jiang,
\newblock Advances in automated treatment planning,
\newblock in {\em Seminars in radiation oncology}, volume~32, pages 343--350, Elsevier, 2022.

\bibitem{li2013automatic}
N.~Li et~al.,
\newblock Automatic treatment plan re-optimization for adaptive radiotherapy guided with the initial plan DVHs,
\newblock Physics in medicine and biology {\bf 58}, 8725--8738 (2013).

\bibitem{gudbjartsson1995rician}
H.~Gudbjartsson and S.~Patz,
\newblock The Rician distribution of noisy MRI data,
\newblock Magnetic resonance in medicine {\bf 34}, 910--914 (1995).

\bibitem{sharma2022deep}
D.~K. Sharma, M.~Chatterjee, G.~Kaur, and S.~Vavilala,
\newblock Deep learning applications for disease diagnosis,
\newblock in {\em Deep learning for medical applications with unique data}, pages 31--51, Elsevier, 2022.

\bibitem{yinghui2025artificial}
W.~Yinghui et~al.,
\newblock Artificial intelligence in four-dimensional imaging for motion management in radiation therapy,
\newblock Artificial Intelligence Review {\bf 58}, 103 (2025).

\bibitem{li2024evaluating}
W.~Li et~al.,
\newblock Evaluating Virtual Contrast-Enhanced Magnetic Resonance Imaging in Nasopharyngeal Carcinoma Radiation Therapy: A Retrospective Analysis for Primary Gross Tumor Delineation,
\newblock International Journal of Radiation Oncology* Biology* Physics {\bf 120}, 1448--1457 (2024).

\bibitem{karimi2021deep}
D.~Karimi, C.~Jaimes, F.~Machado-Rivas, L.~Vasung, S.~Khan, S.~K. Warfield, and A.~Gholipour,
\newblock Deep learning-based parameter estimation in fetal diffusion-weighted MRI,
\newblock Neuroimage {\bf 243}, 118482 (2021).

\bibitem{dai2025r}
Y.~Dai, X.~Jia, Y.-p. Liao, and D.~Jie,
\newblock R-index: A robust metric for IVIM parameter estimation on clinical MRI scanners,
\newblock Magnetic Resonance Imaging , 110560 (2025).

\bibitem{dai2026optimizingimpulsedacquisitionprotocols}
Y.~Dai, X.~Jia, T.~Aguilera, K.~Jiang, A.~P. Rodriguez, I.~Vanhaezebrouck, and J.~Deng,
\newblock Optimizing IMPULSED Acquisition Protocols for Clinical 3T Scanners Through Bayesian Experimental Design, 2026.

\end{thebibliography}



\bibliographystyle{./medphy.bst}    


\end{document}